# Contraction and expansion effects on the substitution-defect properties of thirteen alloying elements in bcc Fe

Wei Liu,[a] Wei-Lu Wang,[a] C. S. Liu,[a,*] Q. F. Fang,[a] Qun-Ying Huang,[b] Yi-Can Wu[b]

[a]*Key Laboratory of Materials Physics, Institute of Solid State Physics, Chinese Academy of Sciences, P. O. Box 1129, Hefei 230031, P. R. China*
[b]*Institute of Plasma Physics, Chinese Academy of Sciences, Hefei 230031, P. R. China*

Proposed as blanket structural materials for fusion power reactors, reduced activation ferritic/martensitic (RAFM) steel undergoes volume expanding and contracting in a cyclic mode under service environment. Particularly, being subjected to significant fluxes of fusion neutrons RAFM steel suffers considerable local volume variations in the radiation damage involved regions. So it is necessary to study the structure properties of the alloying elements in contraction and expansion states. In this paper based on first-principles theory we studied local substitution structures of thirteen alloying elements Al, Co, Cr, Cu, Mn, Mo, Nb, Ni, Si, Ta, Ti, V, and W in bcc Fe and calculated their substitutional energies in the volume variation range from −1.0% to 1.0%. From the structure relaxation results of the first five neighbor shells around the substitutional atom we find that the relaxation in each neighbor shell keeps approximately uniform within the volume variation from −1.0% to 1.0% except those of Mn and the relaxation of the fifth neighbor shell is stronger than that of the third and forth, indicating that the lattice distortion due to the substitution atom is easier to spread in <111> direction than in other direction. The relaxation pattern and intensity are related to the size and electron structure of the substitutional atom. For some alloying elements, such as Mo, Nb, Ni, Ta, Ti and W, the substitutional energy decreases noticeably when the volume increases. Further analysis show that the substitutional energy comprises the energy variation originated from local structure relaxation and the chemical potential difference of the substitutional atom between its elemental crystalline state and the solid solution phase in bcc Fe. We think the approximately uniform relaxation of each neighbor shell around a substitutional atom give rise to a linear decrease in the substitutional energy with the increasing volume.

## 1. INTRODUCTION

For about three decades reduced activation ferritic/martensitic (RAFM) steels have been developed, improved and actively investigated as primary blanket structural materials for fusion power reactors [1-5]. Alloying elements in RAFM steel play an important role in modifying and improving the processing and mechanical properties [6]. No doubt, the evolution of local structures due to alloying elements and the mobility of alloying atoms are directly related to the change of physical and mechanical properties of RAFM. For example, it is believed that the precipitation of Cu atoms in RAFM is one of the main reasons for hardening and embrittlement, which are greatly accelerated under neutron irradiation [7-9]. In a fusion reactor, depending on the pulse lengths, the operating conditions, and the thermal conductivity, oscillating temperature will cause the blanket structural material expanding and contracting in a cyclic mode [10]. Although the thermal expansion coefficient of bcc iron is very small [11], the volume variation may be considerable in the adjacent regions of neutron radiation damages for the existence of lattice deformation. Indeed, our present first-principles calculations of bcc Fe (see section 3) show the local volume change is more than 1.0% for some neighboring crystal cells of a monovacancy or a <110> dumbbell self-interstitial (the most energetically favorable configuration of a self-interstitial atom) in equilibrium state.

In recent years the effects of thermal expansion are usually considered in theoretical investigations on the local structures of foreign interstitials of some small atoms (H, C, and suchlike) and their diffusions in pure metals or alloys [12-14]. These theoretical studies demonstrated that the obtained results are in good agreement with the experimental data when the effect of volume variation is taken into account, indicating that the simulated system is thus close to real conditions. For instance, by using DFT coupled with harmonic transition state theory, Ling and Sholl have studied the absorption and diffusion of carbon in Pd and Pd-based alloys [14]. They have calculated the binding energies and the activation energies of interstitial carbon in octahedral site (O site) and tetrahedral site (T site) in pure Pd and $Pd_{96}M_4$ alloys ($M$ = Ag, Cu and Au) as a function of lattice constant. It should be noted that there is no $M$ atom in the nearest neighbor shell and the next nearest neighbor shell of the carbon atom in these $Pd_{96}M_4$ alloys. They found that in both pure Pd and $Pd_{96}M_4$ alloys, when the lattice constant increases the interstitial carbon becomes stabler and the hopping from O site to T site becomes easier at the same time, and the binding energy and activation energy can be well described by the linear functions of lattice constant. In particular, over a wide temperature region the calculated temperature dependence of carbon diffusivity in pure Pd is in good agreement with the experimental results when the effects of thermal lattice expansion are included. Therefore, an interesting and important question is raised how the volume variation originated from high temperature or neutron radiation damages or external deformation affects the stability and mobility of the alloying atoms and the local structure changes around the alloying elements in RAFM.

In the present work the alloying elements studied include Al, Co, Cr, Cu, Mn, Mo, Nb, Ni, Si, Ta, Ti, V, and W that all form substitution solid solution in RAFM. To provide primary information on the contraction and expansion effect upon the local structures of these alloying elements, we investigate each substitution structure of each alloying element in detail based on first-principles. In addition the substitutional energies of all these alloying elements in contraction and expansion states have been calculated. Our obtained results are helpful for understanding mechanisms of migration and precipitation of these alloying elements in RAFM. This paper is organized as follows. Section 2 describes computational methods and formulae used in this paper. The results of our simulations and the corresponding discussion are reported in Section 3 and 4. Section 3 includes the local structures of a monovacancy and a <110> dumbbell in equilibrium state, and the substitution structures of those alloying elements mentioned above in contraction and expansion states. Section 4 presents the calculated and fitting results of substitutional energies. Finally, a summary is given in section 5.

## 2. METHODOLOGY

The first-principles calculations in this paper are performed using the Vienna *ab initio* simulation package (VASP) that solve Kohn-Sham equation with periodic boundary conditions and a plane-wave basis set [15,16]. Generalized gradient approximation (GGA) formulated by Perdew and Wang (PW91) is used to treat electronic exchange and correlation energies [17,18] with the correlation energy interpolation done by Vosko-Wilk-Nusair method [19]. The pseudopotentials [20,21] are taken from VASP library which are generated within projector augmented wave (PAW) approach that describe transition metals and magnetic systems fairly well [22,23]. For Mo 4p electrons are considered as valence and denoted as "pv". For Nb both the "pv" and another pseudopotential denoted as "sv" that treats 4s and 4p electrons as valence at the same time are used. The standard pseudopotentials are used for all other elements.

There are seven types of modelling boxes used in our calculations, including bcc unit cell, fcc unit cell, hcp unit cell, diamond cubic cell, α-Mn cubic cell [24], cubic supercell I composed of 27 bcc crystal cells, and cubic supercell II composed of 64 bcc crystal cells. In the structure optimization of pure Fe, Mo, Nb, Ta, V and W, bcc unit cells are used; fcc unit cells are used to simulate pure Al, Ni and Cu, and hcp unit cells are used for Co and Ti. For pure Cr, Mn, and Si are antiferromagnetic, antiferromagnetic and diamagnetic, respectively, the cubic supercell I, α-Mn cubic cell and diamond cubic cell are used correspondingly in search of their correct magnetic orders. In the simulation system with a substitution atom the cubic supercell II is used for $Fe_{127}M_1$ (*M* represents all the alloying elements in our study). The cubic supercell II is also used in the system with a monovacancy or a <110> dumbbell. The kinetic energy cutoff is 350 eV and *k*-points sampling is performed using the Monkhorst and Pack scheme for all systems. The *k*-point meshes used for all modelling boxes are listed in table 1. The first-order Methfessel-Paxton method is used for Fermi-surface smearing in all ionic relaxations while the tetrahedron method with Blöchl corrections is used in all accurate energy calculations. The relaxation calculations of these defective systems with a monovacancy or a <110> dumbbell or a substitutional atom are performed using the conjugate gradient algorithm under a constant volume.

The substitutional energy $E_{sub}$ of a foreign atom *M* in bcc Fe matrix is defined as

$$E_{sub} = E(nFe + 1M)_{bcc} - \frac{n}{n+1} \times E((n+1)Fe)_{bcc} - E(M)_{crys}, \tag{1}$$

where $E(nFe + 1M)_{bcc}$ is the energy of a bcc cubic supercell containing *n* Fe atoms and 1 foreign atom, $E((n+1)Fe)_{bcc}$ is the energy of the bcc Fe supercell with the same size as the former, and $E(M)_{crys}$ is the energy per atom of the elemental crystalline *M* in its stablest phase.

In continuum mechanics, based on harmonic solid model, when the volume is changed very slightly, the volume dependence of the system energy can be described by

$$E = E_0 + \frac{1}{2} B_0 \frac{(V - V_0)^2}{V_0}, \tag{2}$$

where $E_0$, $B_0$, and $V_0$ are cohesive energy, bulk modulus and volume in equilibrium state,

respectively. Provided that those tiny solid systems that are comparable with a bcc Fe crystal cell in size can also be described by the above formula, for a system that is composed of $n$ bcc crystal cells containing a substitutional atom, the following formula may be reasonable:

$$E_{sub} = E_{sub0} + c_1(V_{cc} - V_1)^2 + c_2\{[V_{ss} - V_{cc} - (n-1)V_2]^2 - (n^2 - 2n + 2)(V_{pcc} - V_2)^2\}, \quad (3)$$

where $V_{cc}$, $V_{ss}$ and $V_{pcc}$ are the volume of the bcc crystal cell containing a substitutional atom, the volume of the solid system composed of $n$ bcc crystal cells including the substitutional atom, and the volume of a pure bcc Fe crystal cell respectively. In addition, for a certain system with a substitutional atom, $c_1$, $c_2$, $E_{sub0}$, $V_1$ and $V_2$ in formula (3) are all constants that can be determined by fitting of the calculated data. In the structure optimizations of the pure crystals, the energies and volumes are fitted to Murnaghan's equation of state [25]. The fitting of formula (3) is done by using VASP calculated results for all defective systems with a substitutional atom.

## 3. STRUCTURE RESULTS

### 3.1. Equilibrium structures of elemental crystals

Firstly we have calculated the equilibrium lattice constants of all pure crystals, which are summarized in Table 2 along with the reported theoretical and experimental results. Note that all the lattice constants predicted here and the results of the defective system presented below are obtained at 0 K. Although bcc Cr changes into paramagnetism at 300 K [26,27], which is much lower than RAFM's working temperature, we set the system antiferromagnetic by initializing the magnetic moments at each pair of bcc positions (i.e. $(x_1, x_2, x_3)$ and $(x_1 + \frac{1}{2}, x_2 + \frac{1}{2}, x_3 + \frac{1}{2})$) to be antiparallel in our calculations for universal purpose. Pure Mn has a fairly complex phase diagram [28] and in our study it is set to be antiferromagnetic in α-Mn [24] struture. We assume antiparallel moments at each pair of bcc positions to initialize the magnetic structure of pure Mn. The five internal parameters of α-Mn unit cell are fixed to experimental values and only lattice constant is changeable in the optimization. In the magnetism initialization of pure Si we assume collinear spins and set antiparallel moments for each atom at $(x_1, x_2, x_3)$ and its counterpart at $(x_1 + \frac{1}{4}, x_2 + \frac{1}{4}, x_3 + \frac{1}{4})$. As shown in Table 2, it is obvious that our results are in fairly good agreement with the experimental values. Other theoretical lattice constant values listed in Table 2 are all calculated by VASP using PAW-GGA pseudopotentials, and GGA functional of Perdew *et al*. to treat exchange and correlation energies. As can be seen, our calculated results are very close to these reported theoretical values, indicating that our settings are reasonable. It should be noted in search of the equilibrium structure of α-Mn the atom coordinates, volume and shape of the unit cell are optimized simultaneously in Reference 31. It is a common understanding that PAW-GGA pseudopotentials can describe these transition metals and magnetic systems fairly well [22,23,27,30-32]. Furthermore, for Nb and Mo, as 4d elements, have very complicated electron structures, special PAW-GGA pseudopotentials are used for them in this paper. $E(M)$ in formula (1) is calculated by using the optimized equilibrium lattice constants listed in Table 2. The optimized lattice constant of bcc Fe is used in the constant volume calculations with cubic supercell II involving a substitution atom or a monovacancy or a <110> dumbbell.

### 3.2. Local structures around monovacancy and <110> dumbbell

First-principles calculations and experiments both indicate the <110> dumbbell is the stablest self-interstitial in bcc Fe [7]. Therefore, monovacancy and <110> dumbbell are two most important point defects in RAFM. The volumes of nearby crystal cells may be deviated from that in pristine bcc Fe matrix due to the existence of the point defects. For a clear presentation the local structures are described at first. We define the crystal cell that contains a monovacancy or a <110> dumbbell as the central crystal cell. In bcc Fe matrix there are six first nearest group of crystal cells to the defects that each shares one face with the central cell, twelve second nearest group of crystal cells that each shares only one edge with the central cell, and eight third nearest group of crystal cells that each shares only one vertex with the central cell. In Figure 2, if the crystal cell containing a substitutional atom is called the central cell, the crystal cell that contains a second nearest atom to the substitutional atom belongs to the first nearest group, the cell that contains a third nearest atom belongs to the second nearest group and the cell that contains a fifth nearest atom belongs to the third nearest group.

We have investigated the relative volume departure of the central and first three nearest groups of crystal cells from that of pristine cell and our results are presented in Table 3. First we give a description about the results of the monovacancy. According to our calculations, in equilibrium state if the body-centered atom is taken away the volume of the crystal cell will decrease a lot, by 9.73%. For the relaxation is centrosymmetric about the vacancy point, the relative volume departure of all crystal cells in each group are equal. Each crystal cell in the first nearest group shrink by 1.80%, in the second nearest group expands by 0.569% than that in pristine state, and the volume increase is 1.50% for the third nearest group. It should be stressed that the relative volume departure of the third nearest group is in a much greater degree than that of the second nearest group.

Now we turn our attention to the results of <110> dumbbell. In our study after relaxation the distance between the two <110> dumbbell atoms is 1.930 Å, very close to 1.905 Å reported in Reference 7. If an iron atom is put in a crystal cell and a <110> dumbbell is formed, the volume of the cell will increase a great deal, by 17.3%. For any nearby crystal cell, the relative volume departure is dependent on its relative position to the <110> dumbbell. In our present work all these twenty-seven investigated crystal cells compose a cubic supercell and inside the cubic supercell the central crystal cell contains a point defect. Figure 1 is a sketch of the middle layer of the cubic supercell where a <110> dumbbell is located in the central crystal cell. In this layer the four crystal cells of the first nearest group are all swelled considerably, by 6.16%. The volume both decrease by 0.221% for the other two crystal cells of the first nearest group in the upper and lower layer. There are four crystal cells of the second nearest group marked A, C, E and G in this layer, who each shares an edge perpendicular to the dumbbell in the central cell. The volume of crystal cell "A" and "E" both increases by 1.35%, and the relative volume increment is 0.872% for both crystal cell "C" and "G". All other crystal cells of the second nearest group have equivalent volume decrease, i.e. 2.14%. The crystal cells of the third nearest group that each shares a face with crystal cell "A" or "E" all shrink by 5.04%, while all others of this group that each shares a face with crystal cell "C" or "G" swell by 0.514%.

Our calculations indicate that the monovacancy and <110> dumbbell can induce the volume of surrounding crystal cells to expand or shrink considerably from that in pristine state. In the adjacent regions of the much complicated neutron radiation damages in RAFM, such as point defect clusters and dislocation loops, the crystal cells should have considerable volume variations similarly. It is reasonable to think that under neutron irradiation there should be many local regions in RAFM where the crystal cells have considerable volume variations compared with that in pristine state. According to Ling and Sholl's calculations [14], the absorption and diffusion of carbon in Pd and Pd-based alloys are notably influenced by the volume variation in the expansion range of 1.0%. Similarly, the volume variation may influence the stability of the substitutional atom and its migration properties. It is necessary and interesting to investigate the substitution structures and migration properties of these alloying atoms in RAFM in contraction and expansion states. In the following parts, the results of local substitution structures and substitutional energies in contraction and expansion states will be presented and discussed.

### 3.3. Local structures around substitutional atoms

In this subsection we will present the results of local structures around substitutional atoms in contraction and expansion states. In Figure 2 the solid circle marked "s" represents the substitutional atom that is located on the body center of the central crystal cell. Open circles marked "a", "b", "c", "d" and "e" each represents an atom in the first, second, third, forth and fifth nearest shells to the substitutional atom, respectively. As we can see from Figure 2, the second, third and fifth nearest atoms are located on the body center of the first, second and third nearest crystal cells, respectively. There is always a fifth nearest neighbor located just next to the first nearest neighbor in <111> direction.

Figure 3 displays relaxation of the distances from the first five nearest neighbor shells to the substitutional atom in the volume variation range from −1.0% to 1.0%. As we can see, for Nb, the results of two different pseudopotentials, marked "pv" and "sv" are very close to each other, and so are their following results in this paper. There are rather "violent" outward relaxation in all the five neighbor shells around the substitutional atom Mo, Nb, Ta, Ti, and W for they are all much bigger in size than Fe atom. Since the electron structure and atom radius of Co, Cr, Mn, Ni and V are both similar to Fe, there is rather slight relaxation in the neighbor shells around these substitutional atoms. There exists rather strong relaxation in the neighbor shells around substitutional atom Al and Cu. The relaxation of the neighbor shells around Cu atom is in an outward and inward oscillation pattern, which is consistent with the reported results by Domain and Becquart on the first five neighbor shell relaxation around Cu atom in bcc Fe in equilibrium state using VASP software with fully nonlocal Vanderbilt-type ultrasoft pseudopotentials [7]. Since Si is nonmetallic and a little smaller in size than Fe, the relaxation of its neighbor shells are slight and in a particular pattern. The relaxation of each neighbor shell around all substitutional atoms except Mn is approximately uniform within the volume variation from −1.0% to 1.0%. It still needs further study on why the relaxation of the first and fifth neighbor shells around Mn atom change in such a wide range. As we can see in Figure 3, the order of relaxation intensity of the first five neighbor shells is, from strong to weak, the first, second, fifth, third and forth. In bcc lattice the interval between any two neighboring atoms in <111> direction is the shortest among all directions and these neighboring atoms have strong interactions with each other. In the substitution structure the "violent" relaxation of the first neighbor shell in <111> direction strongly acts on the fifth neighbor shell so that its relaxation is stronger than that of the third and forth neighbor shells. Therefore, the lattice distortion originating from the substitution atom in bcc Fe is more noticeable in <111> direction.

Figure 4 displays relative volume departure of the central, the first, second, and third nearest group of crystal cells within the system volume variation from −1.0% to 1.0%. As we can see, the relative volume departure of these crystal cells are rather "large" for atom Al, Mo, Nb,Ta, Ti, and W, and small for atom Co, Cr, Cu, Mn, Ni, Si and V. Although Cr atom is bigger than Fe, the central and the first nearest crystal cells of Cr contracts uniformly within the volume variation from −1.0% to 1.0%, which reflects very good affinity between bcc Fe matrix and atom Cr. The relative volume departure in each neighboring crystal cell group of all substitutional atoms except Mn is approximately uniform within the volume variation from −1.0% to 1.0%. It should be noted that the relative volume departure of the third nearest group of crystal cells is much greater than that of the second nearest group.

## 4. SUBSTITUTIONAL ENERGY RESULTS

For clarity, the substitutional energy curves of these alloying elements in bcc Fe with the volume variation from −1.0% to 1.0% are displayed in Figures 5 (a) and (b). As we can see, the substitutional energy of Mn and Si changes little within the volume variation range from −1.0% to 1.0%, while that of Al, Co, Cr Cu, and V decreases slowly, and for all others it decreases much faster. Since the substitution site is stabler where the substitutional energy is lower, in bcc Fe such an alloying atom will migrate to the swelling zones where its substitutional energy is lower. Provided that the elemental crystalline states of the alloying elements studied in this paper are all the very states into which they will precipitate in bcc Fe, Figure 5 can be used to compare the substitution structure stability for these alloying elements within the volume variation from −1.0% to 1.0%. For example, according to our calculations, in bcc Fe the substitution structure of W is stabler than that of Cu when the volume expands by 0.2%.

In fact, in bcc Fe the substitutional atom is a kind of defect and the substitutional energy calculated in this paper is the defect formation energy. In Reference 33 Zhang and Northrup concluded that the formation energy of a native defect in GaAs can be divided into three parts: the energy variation due to structure relaxation, the chemical potential variation of Ga and As compared with their bulk states, and the electron chemical potential variation derived from free electron transfer. In order to interpret the obtained relations shown in Figure 5 of substitutional energy versus volume variation, following the work by Zhang and Northrup we present a similar formula as follows:

$$E_{sub} = E_{rela} + \mu_s - \mu_{s\,(crys)}.$$

In this formula the chemical potential variations of Fe and free electron are both ignored. Here $E_{rela}$ is the energy variation originated from structure relaxation due to the existence of the substitutional atom. $\mu_s$ is the chemical potential of the substitutional atom in bcc Fe matrix, and $\mu_{s(crys)}$ is its chemical potential in elemental crystalline state. The energy variation $E_{rela}$ is originated from the interaction changes among positive ions in the system since the interaction between positive ions and free electrons can be omitted according to free electron gas model. When the volume is increased, the intervals between any positive ion pairs are lengthened and the corresponding interactions are weakened, so that the energy variation due to the structure relaxation drops. To sum up, in the one-site substitution system of bcc Fe, the substitutional energy includes two parts: the energy variation originated from structure relaxation and the chemical potential difference of the alloying atom between its elemental crystalline state and the solid solution phase in bcc Fe. It is easy to understand that a stronger structure relaxation leads to a larger energy variation. To the best of our knowledge, it remains unknown how the chemical potential of these alloying elements with a solubility of 0.787 at. % in bcc Fe change with volume variation. For an alloying atom in such a dilute solid solution, it sounds plausible that the chemical potential changes only slightly with volume variation from −1.0% to 1.0% because of the little change in both lattice and free electron density.

Provided that the relaxation intensity in each neighbor shell is strictly uniform around a substitutional atom when the volume increases from −1.0% to 1.0%, the energy variation originated from the interaction change among positive ions, as a kind of electrostatic energy, is in linear relation with the relative distance departure in first order approximation. Since the volume is also in linear relation with the relative distance departure in first order approximation when the system volume varies slightly and isotropically, the above-mentioned energy variation should be in linear relation with the relative volume departure. As we can see in Figure 3, the distance relaxation curves of all alloying elements except Mn are approximately even, implying that the relaxation intensity in each neighbor shell is almost uniform in the whole range of volume variation from −1.0% to 1.0%, so it is not surprise to find that except that of Mn all substitutional energy curves in Figure 5 are more or less in a straight line form. Since the relaxation in the neighbor shells around atom Si is rather slight, the energy variation originated from the interaction change among positive ions should play a minor role in the substitutional energy and the chemical potential difference of Si is dominant. As we can see in Figure 5, the substitutional energy curve of Si is very low, which means the chemical potential of atom Si in bcc Fe matrix in the studied system is much lower than in its elemental crystalline state. It is reasonable to think that for those substitutional atoms around whom the relaxation of each neighbor shell is rather strong and even, such as Mo, Nb, Ta, Ti, and W, the descending trend of the substitutional energy curves is the reflection of the decreasing process of the energy variation when the volume is increased from −1.0% to 1.0%.

In Reference 14, Ling and Sholl have fitted carbon's binding energies of O and T sites and O to T activation energy in pure Pd and $Pd_{96}M_4$ alloys using a linear function of the lattice constant from an equilibrium to a swelled state by 1.0%. In our opinion, their fairly good fitting results are due to the rather uniform relaxation intensities of neighbor shells around carbon atom in the volume variation range from 0 to 1.0%. In our present work we fit the calculated substitutional energies using the formula (3) which includes the effect of volume variation of the central crystal cell containing a substitutional atom. As we can see, for most alloying elements the fitting results are very fine, indicating that the expression (3) reasonably describes the volume dependence of the substitutional energy.

In equilibrium state, for Cu, Mn and W who all have very low solubility limits in bcc Fe [9,34], their substitutional energies are rather high compared with others. On the other hand, although Nb, Ta and Ti are very difficult to dissolve in bcc Fe [34,35], their substitutional energies are very low. Furthermore, for Co, Cr and V are full or high soluble in bcc Fe their substitutional energies are not the lowest triplet. In fact the substitutional energies of Co and Cr are much higher than those of Al, Si, Ta and Ti whose solubility limits are not much high. After examining the definition of solution heat, we think substitutional energy equals to the portion of solution heat released or absorbed during the alloying atom enters into bcc Fe matrix and takes a lattice site. So, it can be concluded that the substitutional energy of an alloying element is not directly related to its solubility limit in bcc Fe.

## 5. SUMMARY

In this paper, bcc Fe and the stablest elemental crystalline structures of thirteen alloying elements, i.e., Al, Co, Cr, Cu, Mn, Mo, Nb, Ni, Si, Ta, Ti, V, and W are optimized by fitting VASP calculated lattice constants to Murnaghan's equation of state. We calculated the relative volume departure of the central crystal cell containing a monovacancy or <110> dumbbell and the first, second and third nearest crystal cells in equilibrium state, which indicates the volume variation of adjacent crystal cells are large enough to influence the stability of substitutional atom and its migration. Hence there should be noticeable volume variation for the crystal cells in the adjacent regions to those much complicated neutron radiation damages in RAFM, which is an important motivation for this investigation.

From the structure relaxation results of the first five neighbor shells around the substitutional atom we find that the relaxation in each neighbor shell is approximately uniform within the volume variation from −1.0% to 1.0% except that of Mn. In bcc Fe it is universal that the relaxation of the fifth neighbor shell is stronger than that of the third and forth, which means the lattice distortion originated from the substitution atom is easier to spread in <111> direction than in other direction. The relaxation pattern and intensity are related to the size and electron structure of the substitutional atom. Generally speaking, the relaxation of these neighbor shells around bigger substitutional atoms is stronger and the relaxation around the substitutional atoms that have similar electron structure to Fe is weaker. Relative volume departure of the central and the first three nearest crystal cells of those bigger substitutional atoms is large, while it is small for the substitution atoms whose electron structure is similar to that of Fe.

Our calculations indicate that for some alloying elements, such as Mo, Nb, Ni, Ta, Ti and W, the substitutional energy decreases noticeably when the volume increases from −1.0% to 1.0%, which would influence their migration properties in bcc Fe. We conclude that the substitutional energy comprises the energy variation originated from local structure relaxation and the chemical potential difference of the substitutional atom between its elemental crystalline state and the solid solution phase in bcc Fe. According to our calculations, if the relaxation of each neighbor shell around a substitutional atom are approximately uniform when the system volume increases from −1.0% to 1.0%, the substitutional energy will decline almost linearly with the increasing volume. In this paper the calculated substitutional energies are fitted using a formula which includes the effect of volume variations of the central crystal cell containing the substitutional atom and for most alloying elements the fitting results are very fine.

## ACKNOWLEDGMENT

This work was supported by the Innovation Program of Chinese Academy of Sciences (Grant No.: KJCX2-YW-N35) and the National Magnetic Confinement Fusion Program (Grant No.: 2009GB106005), and by the Center for Computation Science, Hefei Institutes of Physical Sciences.

___________________________________________________

[*]Corresponding author: csliu@issp.ac.cn

Table 1. Converged *k*-point meshes used for all modelling boxes. Herein cubic supercell I is composed of 27 bcc crystal cells and cubic supercell II is composed of 64 bcc crystal cells. bcc, fcc and hcp are acronyms of body-centered cubic, face-centered cubic and hexagonal close-packed, respectively.

| Modelling box | *k*-mesh |
|---|---|
| bcc unit cell | 17×17×17 |
| fcc unit cell | 17×17×17 |
| hcp unit cell | 15×15×13 |
| diamond cubic cell | 11×11×11 |
| α-Mn cubic cell | 5×5×5 |
| cubic supercell I | 5×5×5 |
| cubic supercell II | 3×3×3 |

Table 2. Equilibrium lattice constants of all elemental crystals. PM, FM, AFM and DM are abbreviations of paramagnetic, ferromagnetic, antiferromagnetic and diamagnetic, respectively. Special pseudopotentials used in our calculations are denoted in parentheses.

| Structrure | Lattice constant (Å) | | |
|---|---|---|---|
| | This paper | Expt.[a] | Other calculated results |
| Al fcc PM | 4.048 | 4.05 | 4.048[f] |
| Co hcp FM | a=2.496 c=4.024 | a=2.51 c=4.07 | a=2.494 c=4.025[f] |
| Cr bcc AFM | 2.849 | 2.88 | 2.85[d] 2.850[e] 2.847[f] |
| Cu fcc PM | 3.636 | 3.61 | 3.631[f] |
| Fe bcc FM | 2.833 | 2.87 | 2.83[c,d] 2.831[e] 2.822[f] |
| Mn αMn cube AFM | 8.557 | 8.89[b] | 8.668[g] |
| Mo bcc PM | 3.173 (pv) | 3.15 | 3.178[f] |
| Nb bcc PM | 3.323 (pv) 3.310 (sv) | 3.30 | 3.322[f] |
| Ni fcc FM | 3.521 | 3.52 | 3.517[f] |
| Si diamond cube DM | 5.468 | 5.43 | 5.461[h] |
| Ta bcc PM | 3.310 | 3.31 | 3.320[f] |
| Ti hcp PM | a=2.915 c=4.623 | a=2.95 c=4.69 | a=2.929 c=4.628[f] |
| V bcc PM | 2.977 | 3.02 | 2.992[f] |
| W bcc PM | 3.176 | 3.16 | 3.190[f] |

[a]All experimental results are taken from reference 28, except that of Mn.
[b]Reference 24.  [f]Reference 30.
[c]Reference 22.  [g]Reference 31.
[d]Reference 27.  [h]Reference 32.
[e]Reference 23.

Table 3. Relative volume departure of the crystal cells surrounding the point defects in equilibrium state.

| Defect | Relative volume departure from pristine crystal cell (%) | | | |
|---|---|---|---|---|
| | Cental cell | Cell in first group | Cell in second group | Cell in third group |
| Vacancy | -9.73 | -1.80 | 0.569 | 1.50 |
| <110> | 17.3 | -0.221, 6.16 | -2.14, 1.35, 0.872 | -5.04, 0.514 |

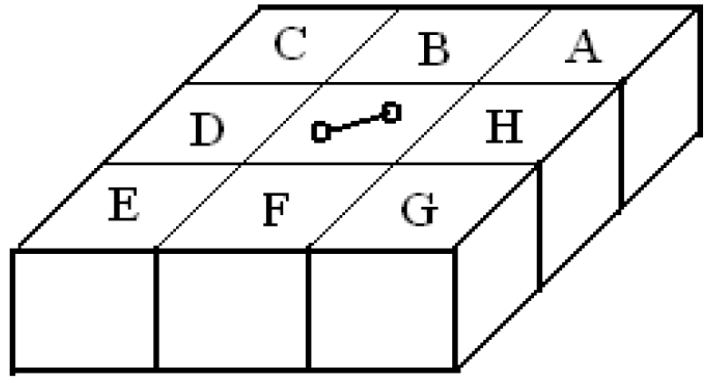

Figure 1. Sketch of the middle layer of the cubic supercell composed of twenty-seven bcc Fe crystal cells with a <110> dumbbell in the central cell.

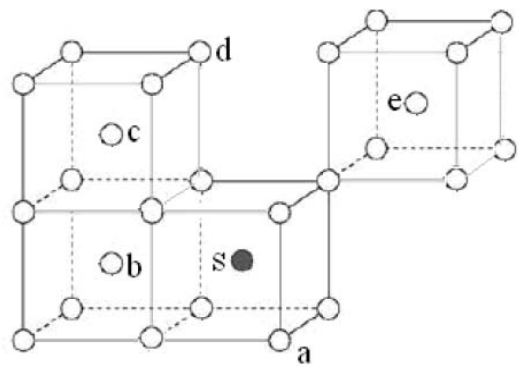

Figure 2. First five nearest atoms to the substitutional atom. Solid circle marked "s" is the substitutional atom. Empty circles marked "a", "b", "c", "d" and "e" are the first, second, third, forth and fifth nearest atoms to the substitutional atom, respectively.

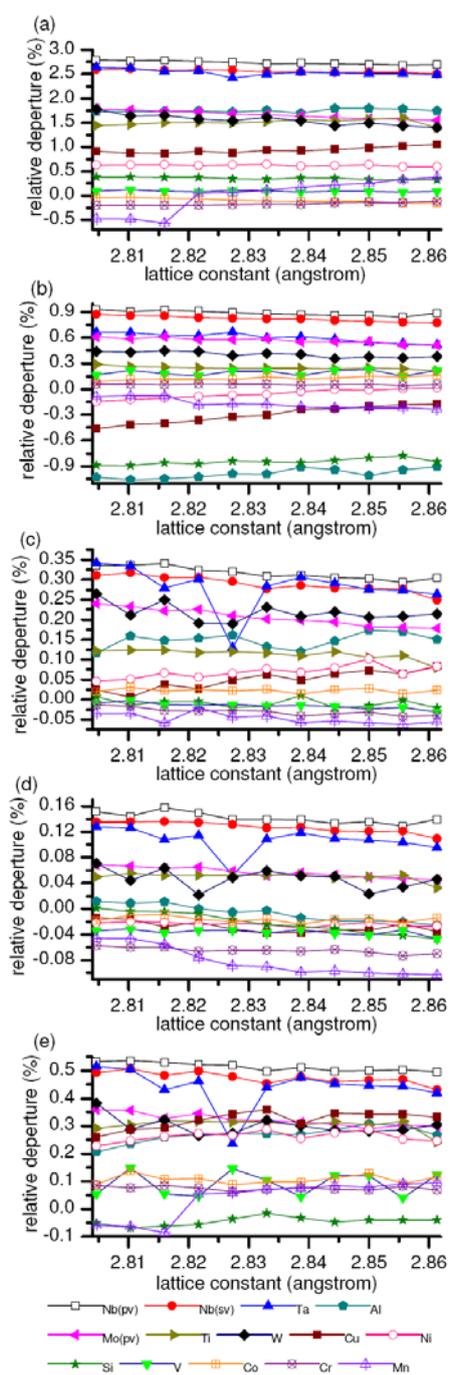

Figure 3. (Color online). Relative distance departure from the first five nearest neighbor shells to the substitution atom: (a) the first nearest neighbor shell, (b) the second nearest neighbor shell, (c) the third nearest neighbor shell, (d) the forth nearest neighbor shell, and (e) the fifth nearest neighbor shell.

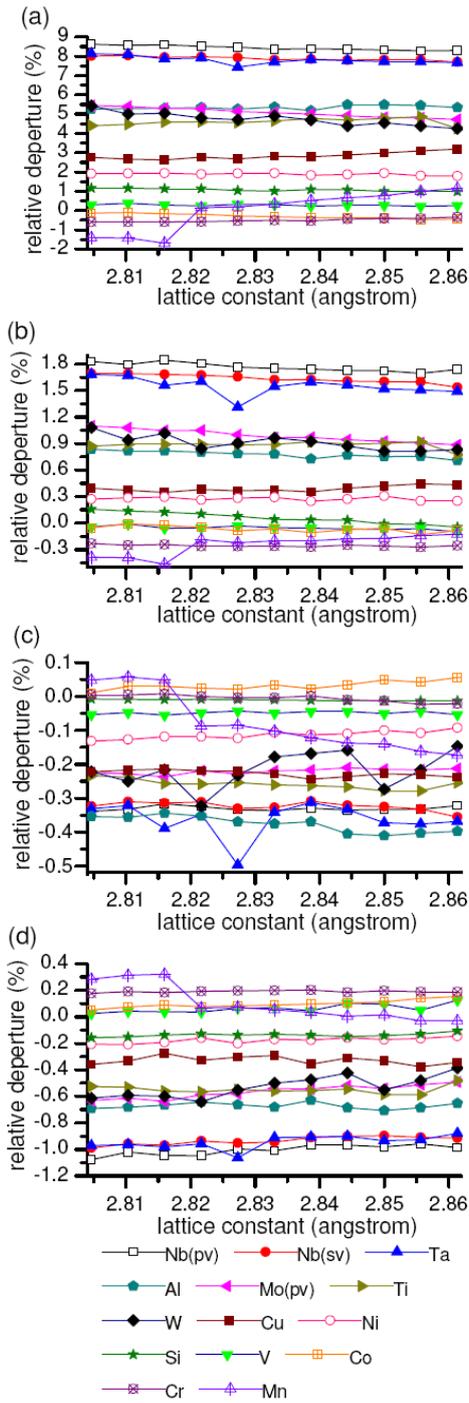

Figure 4. (Color online). Relative crystal-cell-volume departure from pristine state due to the existence of substitutional atoms: (a) the central cell, (b) the first nearest neighbor cells, (c) the second nearest neighbor cells, and (d) the forth nearest neighbor cells.

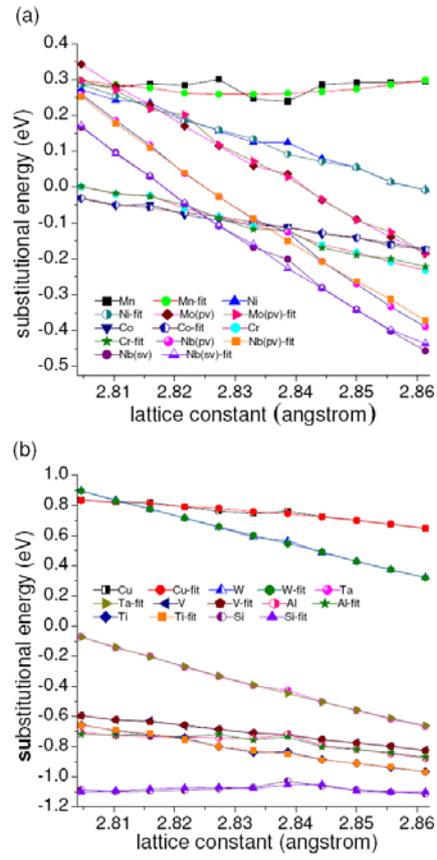

Figure 5. (Color online). Calculated and fitted substitutional energy curve